\documentclass[aps,prl,reprint,showpacs]{revtex4-1}

\usepackage{graphicx}
\usepackage{dcolumn}
\usepackage{bm}
\usepackage{epsfig}
\usepackage{amsmath}
\usepackage[usenames,dvipsnames]{color}
\usepackage{booktabs}

\begin{document}

\author{Mahmoud M. Asmar}
\email{asmar@phy.ohiou.edu} \affiliation{Department of Physics and
Astronomy and Nanoscale and Quantum Phenomena Institute, Ohio
University, Athens, Ohio 45701-2979, USA}
\author{Sergio E. Ulloa}
\email{ulloa@ohio.edu} \affiliation{Department of Physics and
Astronomy and Nanoscale and Quantum Phenomena Institute, Ohio
University, Athens, Ohio 45701-2979, USA}

\title{Rashba Spin Orbit Interaction and Birefringent Electron Optics in
Graphene}

\begin{abstract}
Electron optics exploits the analogies between rays in geometrical
optics and electron trajectories, leading to interesting insights
and potential applications. Graphene, with its two-dimensionality
and photon-like behavior of its charge carriers, is the perfect
candidate for the exploitation of electron optics. We show that a
circular gate- or doping-controlled region in the presence of Rashba
spin-orbit interaction in graphene may indeed behave as a Veselago
electronic lens but with two different indices of refraction. We
demonstrate that this birefringence results in complex caustics
patterns for a circular gate, selective focusing of different spins,
and the possible direct measurement of the Rashba coupling strength
in scanning probe experiments.
\end{abstract}

\pacs{ 71.70.Ej, 75.76.+j, 72.10.Fk, 42.25.Lc }

\maketitle

The analogies between geometrical optics and electron trajectories
have resulted in a number of interesting proposals for device
applications \cite{Baldwin}, where interfaces play a similar role to
that played by transparent interfaces in physical optics.  This
leads to the manipulation and control of electron trajectories,
where the major factor determining the electron dynamics is the
change in group velocity through these interfaces, mimicking
refringent physical optics and lenses.  Such change in group
velocity is typically achieved by local gating, which modulates
carrier densities and fixes the corresponding index of refraction.

Optical birefringence in materials results from crystal anisotropies
which are manifested as different group velocities for different
polarizations of the propagating light in the material. In this
paper, we show that an equivalent phenomenon to optical
birefringence in electron optics is feasible in two dimensional
graphene, which in essence reflects the intrinsic crystal structure even at
large electronic wavelengths.  The effect
requires the presence of Rashba spin-orbit interaction, where the
different group velocities depend on the chirality of the electronic
states, mimicking the light polarization dependence of the
group velocities in optical birefringent materials.

The low energy dispersion of electrons in graphene is centered near
two inequivalent points in the Brillouin zone, the $K$ and $K'$ or
Dirac points \cite{novo,elctronincproperties,seminov}. The
``massless'' nature of electrons results in novel phenomena such as
the Klein paradox \cite{Klein,gate3,elctronincproperties}, which
leads to full transparency of a sharp gated interface for normal
incident electrons, and a high probability of transmission for
incoming electrons with finite angles. The linear dispersion of
electrons is also evocative of photons, prompting a number of
proposals and experiments to probe optical analogs with charge
carriers \cite{wvaegides}, aided in great measure by the high
electron mobility in this unique material. In fact, the transparency
of barriers and ability to gate regions of the system to change the
sign of carriers can lead to the use of graphene gate-controlled
interfaces as electronic lenses that follow Snell's law with
negative index of refraction and allow the implementation of
electronic analogues of Veselago optics
\cite{Veselago,caustics2,caustics}.

The study of spin transport properties of suspended and deposited
graphene is rapidly becoming an important area of research.
Experiments have achieved spin polarized injection of electrons and
measure spin valve effects \cite{valves1,valves2,valves}, spin
polarized currents with long coherence lengths in suspended graphene
\cite{coherence}, and shorter coherence lengths for deposited
samples or samples containing impurities that enhance spin orbit
effects \cite{charged,experiment1}, such as hydrogen or gold
\cite{golddep,impurityso}. An important ingredient determining the
carrier spin dynamics in this and other materials is the spin orbit
interaction (SOI). The \emph{intrinsic} SOI respects all the lattice
symmetries in graphene and results in a small energy gap at the
Dirac points \cite{quntum}. The \emph{extrinsic} or Rashba SOI
results from the lack of inversion symmetry due to perpendicular
electric fields, substrate effects, chemical doping, or curvature of
graphene corrugations \cite{curvature2,Ni1}. In this paper we focus
on the effects of the Rashba SOI in the spin dynamics of carriers
away from the neutrality point as the spin-orbit interaction can be
controlled by external factors \cite{intrinsicstr}.

\begin{figure*}
\includegraphics[scale=0.65]{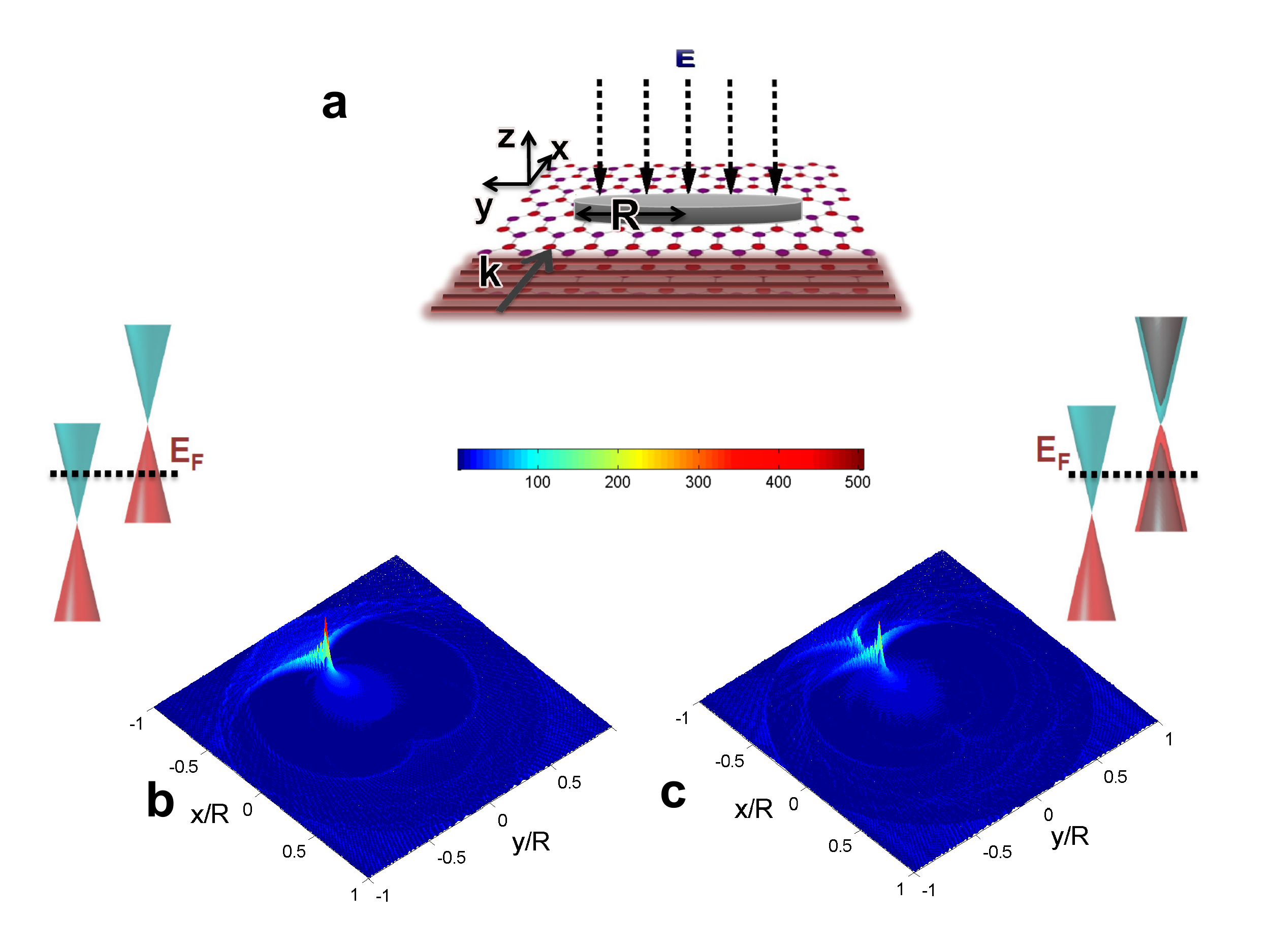}
\caption {(Color online) \textbf{a}) Electron flux in graphene
coming along the $x$-direction onto a circular gate potential
covering an area of radius $R$; the normal electric field reverses
the carrier character from electron to hole and also generates a
Rashba spin-orbit field. \textbf{b}) Three dimensional map of
$|\psi|^{2}$ near and inside gated region centered at $(x,y)=(0,0)$
(normalized to incident flux). The electronic patterns produced
inside the gated region are analogous to those produced by the
refraction of light in a circular medium with refractive index
$n=-1$ (Veselago lens). The electronic index of refraction is
determined by the ratio of wave numbers in and out of the gated
obstacle: both regions are characterized by a linear electronic
dispersion (upper left inset), and in this case, $kR=300$ (electron)
and $k'R=300$ (hole). High intensity maxima correspond to caustics
with $p=1$ $(x<0)$ and $p=2$ $(x>0)$, where $p-1$ is the number of
internal reflections. $x$ and $y$ coordinates in units of the radius
$R$. \textbf{c}) In the presence of Rashba spin orbit interaction
the electronic dispersion in the gated region is modified, so that
scattering particles have access to two different wave numbers
(upper right inset). Scattering produces electronic patterns as
shown in the three dimensional map $|\psi|^2$, with $kR=300$,
$VR/\hbar v_{F}=600$, $\lambda_{R}R/\hbar v_{F}=100$, for
\emph{$\uparrow$-spin incoming flux}. The $p=1$ and $p=2$ caustics
and cusps are doubled, associated with states of different chirality
in graphene.  The pattern can be described by the refraction from a
medium with two (negative) indices, $n_\pm$. Here, the degree of
birefringence is $\Delta n=n_{-}-n_{+}=0.71$. The dual pattern
persists even for spin-unpolarized incidence. \label{fig1}}
\end{figure*}

To study these effects, we consider a graphene sheet with a circular
gate potential (or corresponding doping/intercalation profile)
covering an area of radius $R$, where the electric field (or doping)
reverses the carrier character from electron to hole and also
generates a Rashba SOI of strength $\lambda_{R}$, as shown in
Figure~\ref{fig1}a. Electron scattering from such gated regions has
been shown to result in pronounced wave function intensity maxima
due the constructive interference between different partial wave
components \cite{caustics2}. These patterns are similar to the
optical caustics which develop by light refraction through a shaped
medium, and belong to a class of cusps in catastrophe theory
\cite{beerycustics}. We should comment that recent experiments have
demonstrated gate tunability, which makes the study of electron
optics on graphene viable and controllable \cite{gate,gate1,gate3}.
Moreover, imaging of the patterns resulting from the electron flow
through these nanoscale structures can be achieved through STM
\cite{stm,stmM} or other scanning probe techniques
\cite{sgate2,sgate3}.

The Hamiltonian of the system has the form $H=H_{o}+H_{V}+H_{R}$,
where $H_{o}/\hbar v_F=\sigma_{x}k_{x}+\sigma_{y}k_{y}$ describes
free electrons in graphene with momentum $\vec{k}=(k_{x},k_{y})$
away from the $K$ point. $H_{V}=V\vartheta(R-r)$ represents the
gated or doped region of strength $V$ and
$H_{R}=\lambda_{R}(\sigma_{x}s_{y}-s_{x}\sigma_{y})\vartheta(R-r)$
is the Rashba SOI in this circular region for $\vec{k}$ near the $K$
point \cite{seminov,quntum}, where
$\{\sigma_{\mu}\}$ and $\{s_{\mu}\}$ are Pauli matrices representing
the electron pseudospin $(A,B)$ and spin $(\uparrow,\downarrow)$,
respectively, $\lambda_{R}$ is the strength of the Rashba
interaction, and $\vartheta$ is the Heaviside function determining
the radius $R$ of the region. The Hamiltonian is parameterized in
terms of $\hbar v_{F}$, where $v_{F}=10^{6} m/s$ is the carrier
Fermi velocity \cite{novo,elctronincproperties}. We are interested
in Rashba regions with $R\gg a$, where $a$ is the lattice constant
of graphene, in order to adopt the continuum description of
graphene.

The cylindrical symmetry of the system allows one to write the
eigenfunctions in terms of four linear differential equations
coupling the different spinor components,
$\mathbf{\psi}_{j}=(\psi_{A\uparrow},\psi_{B\uparrow},\psi_{A\downarrow},\psi_{B\downarrow})^{T}$.
The symmetries result in the conservation of the $z$-component of
the total angular momentum in the system, $\left[J_{z},H\right]=0$,
where $J_{z}=L_{z}+\hbar s_{z}/2+\hbar\sigma_{z}/2$, and
$L_{z}=-i\hbar\partial_{\theta}$ is the orbital angular momentum; as
such, the eigenstates describing the system can be labeled according
to their total angular momentum, $j$, with
$J_{z}\mathbf{\psi}_{j}=j\hbar\mathbf{\psi}_{j}$ (see Supplemental
Material).

The scattering of an incident flux of electrons on the gated
obstacle is studied through a spin-dependent generalization of the
partial wave component method \cite{elestic}. This utilizes the
analytical solutions for the spinors together with their asymptotic
expansions and boundary conditions in order to obtain the scattering
amplitudes. The latter allow the direct evaluation of wave functions
inside and outside the Rashba region, depending on the incident
energy, momentum and spin content of the incident flux, as well as
system parameters (see Supplemental Material). The momentum outside
the scattering region is given by $k=E/\hbar v_{F}$, while the
momentum inside takes \emph{two different values},
$k_{\pm}=\sqrt{\left(E-V\right)^{2}\pm2\lambda_{R}(E-V)}/\hbar
v_{F}$, for the given energy $E$, due to the presence of the Rashba
SOI (see Figure~\ref{fig1}).

Formation of caustics has been described in the absence of SOI
\cite{caustics}, resulting in beautiful scattering patterns. The
caustics and cusps can be understood in terms of geometrical optics,
and characterized by a negative index of refraction
$n=-|E-V|/|E|=-k'/k$, where $k'$ is the wave number inside the gated
region \cite{caustics2} (Veselago lensing \cite{Veselago}). For an
incident electron flux along the $x$-direction, the position of the
cusps produced inside a circular gated region can be shown to be
given by $x_{cusp}=(-1)^{p}/(|n|-1+2p)$, for $p-1$ internal
reflections of the ray inside the region \cite{caustics} (see
Figure~\ref{fig1}b).

In contrast, in a system with Rashba SOI the two wave numbers for a
given energy allow for two refraction indices, associated with the
two chiral solutions of the Dirac equation, $n_{\pm}=-k_{\pm}/k$.
Correspondingly, the optical analogue results in two different cusp
locations (see Fig.\ \ref{fig1}c) given by
\begin{equation} \label{eq1}
 x^{\pm}_{cusp}= \frac{(-1)^{p}}{|n_{\pm}|-1+2p}\; .
\end{equation}
As we will see below, this simple description is borne out by the
full quantum calculation of the scattering patterns. We should
emphasize that despite the optical analogy, birefringence here has a
clear quantum mechanical origin, as the group velocity and built in
phases of the different chiral states are different due to the
presence of the Rashba interaction.

\begin{center}
\begin{figure}
\includegraphics[scale=0.45]{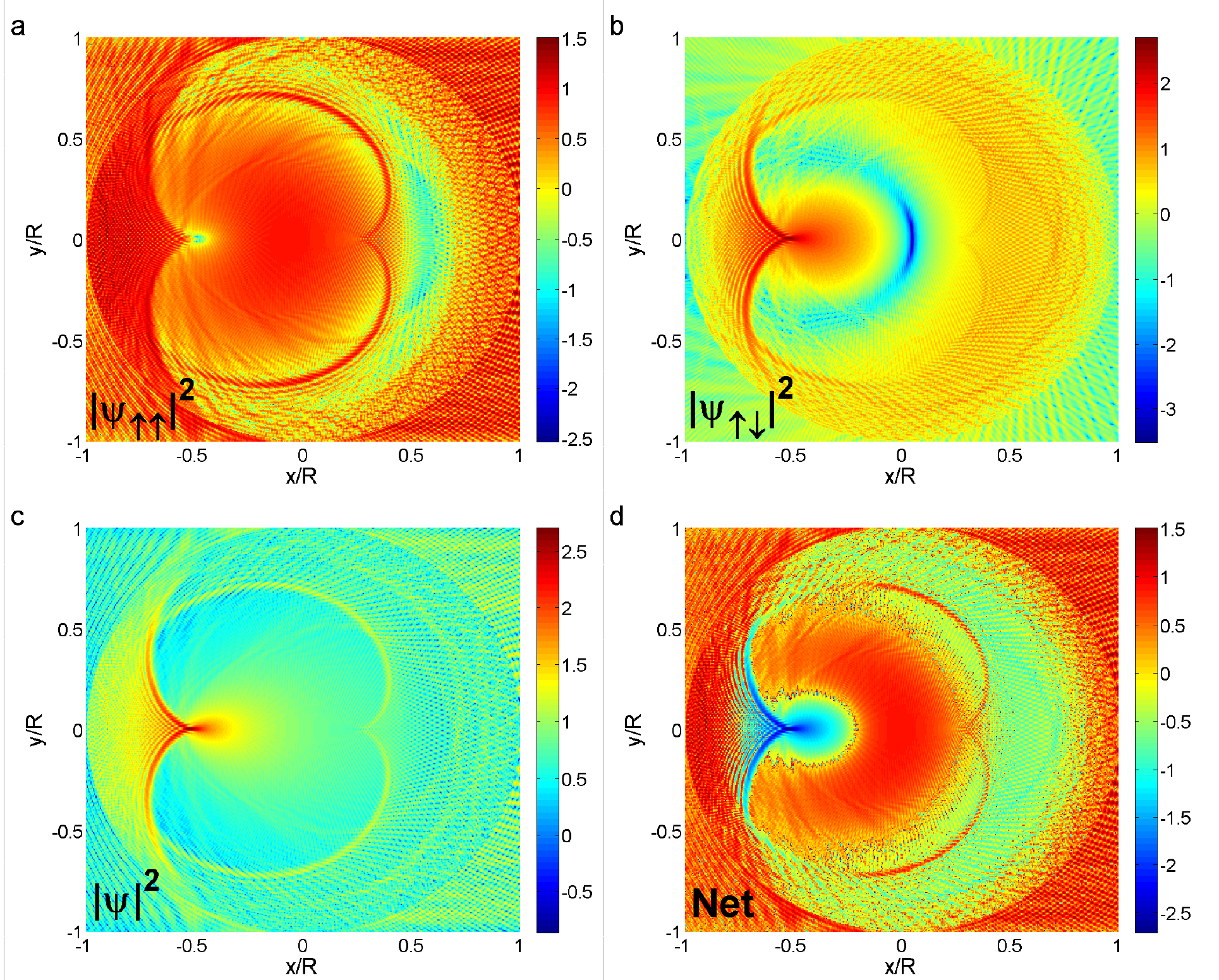}
\caption {(Color online) Probability density patterns (scale bars
show log of amplitudes, normalized to incident flux) resulting from the scattering of
\emph{incoming $\uparrow$-spin} electron wave along the
$x$-direction with $kR=300$, in the presence of a gate potential
$VR/\hbar v_{F}=600$, and Rashba coupling $\lambda_{R}R/\hbar
v_{F}=3$. This arrangement leads to a system with  a small degree of
birefringence $\Delta n=n_{-}-n_{+}=0.02.$ \textbf{a}) The
spin-preserving component $|\psi_{\uparrow\uparrow}|^{2}$ and
\textbf{b}) spin-flip component $|\psi_{\uparrow\downarrow}|^{2}$
display Rashba oscillations with a wavelength covering a large
region of the scattering target, due to the small $\lambda_{R}$.
\textbf{c}) Total wave function
$|\psi|^{2}=|\psi_{\uparrow\uparrow}|^{2}+|\psi_{\uparrow\downarrow}|^{2}$,
displays caustics and cusps that remain almost unchanged from the
case in which the Rashba interaction is absent. \textbf{d}) Net spin
$\eta\sim|\psi_{\uparrow\uparrow}|^{2}-|\psi_{\uparrow\downarrow}|^{2}$
quantifies the predominance of a given spin in different regions of
the system. The large wavelength of Rashba oscillations leads to a
cusp at $x\approx-0.5$ with a net $\downarrow$-spin (blue) while the
annular region near $x\simeq0$ has a net $\uparrow$-spin (red).
 \label{fig2}}
\end{figure}
\end{center}

The presence of the Rashba interaction causes the spin of the
incoming electron to precess as it travels along the scattering
region; this spin precession is seen as oscillations in the
amplitudes of the spin components of the wave function (Rashba
oscillations), where a clear spatial frequency can be identified for
different $\lambda_{R}$ values: shorter precession wavelength for
larger $\lambda_{R}$. For small Rashba coupling, the wavelength of
these oscillations is comparable to the size of the gated region,
leading to the presence of different spins over large areas of the
scattering disk. Figure \ref{fig2}a,b shows the wave functions
inside and outside the scattering region for spin up and down
components for $\lambda_{R}R/\hbar v_{F}=3$; take notice these are
results {\em for spin-up incidence}. Notice that both spin
components display caustics inside the scattering region. Moreover,
one observes that the wave function around the caustics has a net
spin content, as shown by $\eta={\rm
sgn}(\Delta)|\log_{10}(|\Delta|)|$, where
$\Delta=|\psi_{\uparrow\uparrow}|^2-|\psi_{\uparrow\downarrow}|^2$,
in Figure~\ref{fig2}d. For this set of parameters, the $p=1$ caustic
$(x<0)$ is predominantly spin-down, while the $p=2$ caustic $(x>0)$
is predominantly spin-up. The total wave function,
$|\psi|^2=|\psi_{\uparrow\uparrow}|^2+|\psi_{\uparrow\downarrow}|^2$
shows no major changes due to the Rashba SOI, since $\lambda_{R}$ is
small and the two different wave numbers are nearly equal, $k_{-}-
k_{+}\approx 2\lambda_{R}/\hbar v_{F}$; see Figure~\ref{fig2}c.
Correspondingly, the two indices are also very similar,
$n_{-}-n_{+}=0.02$, and  the pattern of caustics and cusps is
essentially unchanged from the $\lambda_{R}=0$ case.

\begin{center}
\begin{figure}
\includegraphics[scale=0.45]{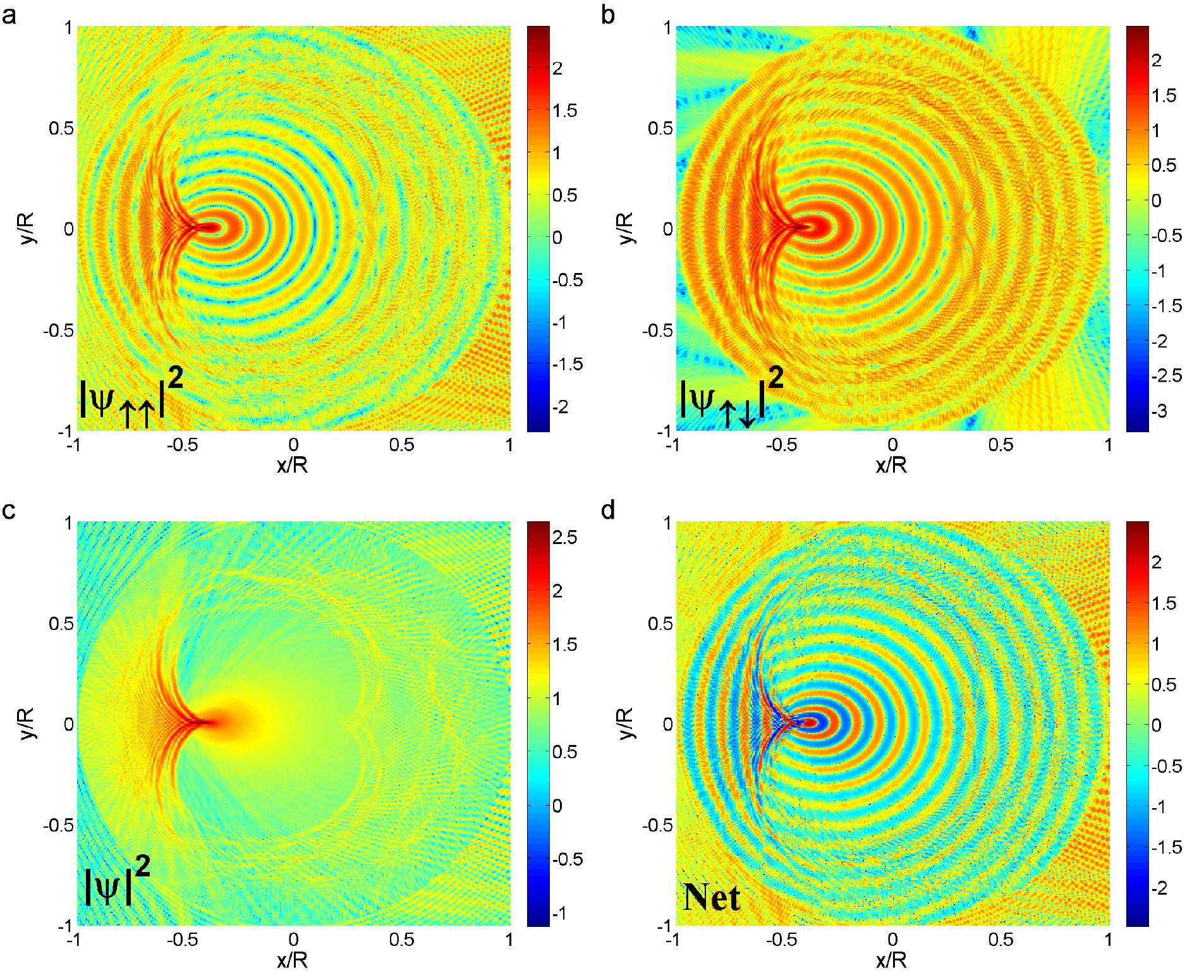}
\caption {(Color online) Wave function maps (scale bar show log of
amplitudes; normalization to incident flux) similar to those in
Figure~\ref{fig2} but for larger $\lambda_R$. Here, $kR=250$,
$VR/\hbar v_{F}=600$ and $\lambda_{R}R/\hbar v_{F}=30$. This
arrangement leads to a system with a degree of birefringence $\Delta
n= n_{-}-n_{+}=0.24$. The different energy of the incoming wave
leads to modified positions of caustics and cusps. \textbf{a})
$|\psi_{\uparrow\uparrow}|^{2}$ and \textbf{b})
$|\psi_{\uparrow\downarrow}|^{2}$ show clearer and sharper Rashba
oscillations with shorter wavelength. Notice that the number of
oscillations inside the gated region is approximately
$\lambda_{R}R/2\hbar v_{F}=15$ for both spins, as shown by the rings
around the main cusp. \textbf{c}) The total wave function,
$|\psi|^{2}$, displays clear duplicate sets of caustics and cusps
manifesting the birefringent character of the scattering, due to the
large Rashba interaction and associated different group velocities.
\textbf{d}) The net spin $\eta$, shows variation of the two spin
components along the the different cusps and caustics. \label{fig3}}
\end{figure}
\end{center}

For larger values of the Rashba interaction the oscillations become
even sharper, with shorter wavelength, and their number inside the
gated region is approximately $\lambda_{R}R/2\hbar v_{F}$ for both
spins. The larger value of $\lambda_{R}$ leads to two very different
wave numbers inside the scattering region, $k_{\pm}$, different
group velocities for the charge carriers, and different indices
$n_{\pm}$. In the case of $\lambda_{R}R/\hbar v_{F}=30$,
Figure~\ref{fig3}, we notice a clear doubling of the cusps,
especially evident in panel c for the total wave function. Moreover,
the cusp positions are in agreement with the anticipated values
shown by Eq.\ \ref{eq1}, as one can easily verify.

Let us comment on the system conditions for the observation of these
effects. We have first assumed a semi-classical limit, with $kR\gg1$
and $k_{\pm}R\gg1$; second, we have assumed the absence of
intervalley scattering ($K$ to $K'$), which requires that the
characteristic length of potential change must be much larger than
the lattice constant. These two rather sensible requirements
introduce constraints on the values of the Fermi energy of the
incoming electron flux, the gate voltage, and the size of the gated
region. Recent experiments have reported values of the Rashba
interaction in the range $10$-$100meV$ for graphene samples grown on
Ni and intercalated with Au atoms \cite{Ni1}.  This coupling is 2-3
orders larger than the spatially random spin orbit generated by
intrinsic ripples in graphene \cite{curvature2}. Assuming possible
values of these parameters in experiments as $E=80meV$, corresponding
to carrier concentration $n_c \simeq 0.5 \times 10^{12}cm^{-2}$,
$V=200meV$, and $\lambda_{R}=10meV$, would
require $R\geq1000nm$ and lead to $n_{-}-n_{+}\geq0.24$, allowing
the observation of the patterns shown in Figure~\ref{fig3}c, and its
clear detection in scanning probe experiments, as
$|x^{+}_{cusp}-x^{-}_{cusp}|\ge42nm$. We should also notice that
although we have used spin-polarized injection in our calculations,
to expose the spin character of the caustic pattern, this is {\em
not} a necessary ingredient to observe the birefringence.
Unpolarized electron injection will obviously erase the net spin
structures we have discussed. However, the appearance of duplicate
caustics and cusps would persist and result in an \emph{identical}
spatial structure of the total wave function to the ones shown
above. Similarly, $K$-$K'$ coupling would preserve the duplicate
caustic pattern, although with decreased contrast, as the enhanced
backscattering would reduce the overall transparency of the region.

In conclusion, we studied the effects of Rashba SOI on the
scattering of electrons in graphene. For spin polarized injection,
we find a spatial modulation of the carrier spin, with
characteristics that depend on the strength of the Rashba
interaction and may be important in the implementation of spin based
devices in graphene systems. More important, we find the selective
formation of spinful cusps and caustics in the region containing
large Rashba SOI, opening the possibility of spin beam filters and
splitters. Moreover, the manifestation of electronic birefringence
at relatively large but experimentally achievable values of the
Rashba SOI \cite{Ni1}, is the doubling of caustics and cups produced
by these refringent Veselago lenses; the spacing between the two
different chiral cusps is proportional to the strength of the Rashba
interaction in the system.  As such, the birefringence can be
used to determine the strength of the Rashba SOI in the region of
interest in the system.

We believe that the concept of birefringent electron optics
established here can lead to important consequences in experiments
and applications.  For example, using guiding gates \cite{wvaegides}
over graphene samples grown on Ni[111], or creating a suitable
spatial pattern of gold intercalation, one could design a
birefringent waveguide with two different critical angles
($\theta_{\pm}= \sin^{-1}(k_{\pm}/k)$) for the two orthogonal
chiral states of the system, leading to the independent propagation
of chiral states, with the corresponding spatial modulation of the
spin componenets \cite{spinpump}.

We thank M. Zarea and N. Sandler for useful discussions and the
support of NSF CIAM/MWN and PIRE grants.

\end{document}